\documentclass[a4paper,fleqn,usenatbib]{mnras}

\usepackage{newtxtext,newtxmath}

\usepackage[T1]{fontenc}
\usepackage{ae,aecompl}

\usepackage{graphicx}	
\usepackage{amsmath}	

\newcommand{\Msun}{\mbox{$M_{\odot}$}}
\newcommand{\Rsun}{\mbox{$R_{\odot}$}}
\newcommand{\Lsun}{\mbox{$L_{\odot}$}}
\newcommand{\kms}{\mbox{km s$^{-1}$}}

\title[Eclipsing binaries in open cluster NGC\,3532]{Two newly identified eclipsing binaries in open cluster NGC\,3532}

\author[\"Ozdarcan O.]{
\"Ozdarcan O.$^{1,2}$\thanks{E-mail: orkun.ozdarcan@ege.edu.tr}
\\
$^{1}$Ege University, Science Faculty, Department of Astronomy and Space Sciences, 
35100 Bornova, \.{I}zmir, Turkey\\
$^{2}$T\"UB\.ITAK National Observatory, Akdeniz University Campus, 07058 Konyaalt\i, Antalya, Turkey\\
}

\date{Accepted XXX. Received YYY; in original form ZZZ}

\pubyear{2020}

\begin{document}
\label{firstpage}
\pagerange{\pageref{firstpage}--\pageref{lastpage}}
\maketitle

\begin{abstract}
We present light curve analyses of two newly identified detached eclipsing binaries,
HD\,96609 and HD\,303734, in the region of the richly populated open cluster NGC\,3532. 
HD\,96609 is composed of two main sequence stars (B9-A0V + A2V) with masses and radii of 
$M_{1}=2.66\pm0.02$\Msun, $M_{2}=1.84\pm0.01$\Msun, $R_{1}=2.740\pm0.006$\Rsun, 
$R_{2}=1.697\pm0.005$\Rsun. The positions of the components on $log~M-log~R$ plane suggests 
log(age/yr) 8.55, corresponding $350\pm40$ Myr of age, which agrees with the $300\pm100$ 
Myr age of NGC\,3532 estimated in previous studies. We find the distance of HD\,96609 as 
$460\pm17$ pc, which is consistent with the $484^{+35}_{-30}$ pc distance of NGC\,3532, 
estimated from GAIA parallaxes. HD\,303734 is an interesting totally eclipsing binary with 
a quite shallow secondary eclipse. Using photometric properties of the system in conjunction 
with theoretical calibrations, we estimate that HD\,303734 consists of A6V + K3V components. 
HD\,96609 and HD\,303734 are the second and third eclipsing binaries discovered in the 
region of NGC\,3532, after the first one, GV\,Car.

\end{abstract}




\section{Introduction}\label{sec_intro}
NGC\,3532 is one of the most spectacular and richly populated open cluster located in the southern 
sky. Although these properties, less number of studies exist on the cluster compared to other 
well-known southern open clusters. The first comprehensive $UBV$ photoelectric photometry of the 
cluster was presented by \citet{koelbloed_UBV_1959BAN....14..265K}, who used the main-sequence 
fitting method and determined the distance of the cluster as $432\pm40$ pc with an $E(B-V)$ value 
of 0\fm01 and an approximate age of 100 Myr. Subsequent broad-band photometric studies were 
provided by \citet{fernandez_salgado_UBV_1980A&AS...39...11F}, \citet{wizinowich_UBV_1982AJ.....87.1390W} 
and \citet{claria_lapasset_UBV_1988MNRAS.235.1129C} with additional DDO and Washington photometry. 
Moreover, Str\"omgren $uvbyH_{\beta}$ photometry of the cluster were presented by 
\citet{eggen_stromgren_1981ApJ...246..817E} and \citet{schneider_stromgren_1987A&AS...71..147S}. 
The most recent, comprehensive and precise $BVRI$ photometry of the cluster, going deeper 
magnitudes of $V=22^{m}$, was published by \citet{clem_landolt_2011AJ....141..115C}, who revised the
distance and $E(B-V)$ value of the cluster as $492^{+12}_{-11}$ pc and $0\fm028\pm0\fm006$, 
respectively. They also revised the age of the cluster as $300\pm100$ Myr by using fitting 
overshooting isochrones to the upper main sequence of the cluster in colour-magnitude diagram. 
The distance estimation of \citet{clem_landolt_2011AJ....141..115C} matches the most recently reported
distance of the cluster, $484^{+35}_{-30}$ pc, which is computed via parallax measurements included in 
the second GAIA data release \citep{strassmeier_ngc3532_2019A&A...622A.110F, 
gaia_main_2016_2016A&A...595A...1G, gaia_dr2_2018A&A...616A...1G}. Recent studies indicate that
the cluster possesses near solar metallicity \citep{bossini_et_al_2019A&A...623A.108B, 
strassmeier_ngc3532_2019A&A...622A.110F}.

\begin{figure}
	\includegraphics[scale=0.9]{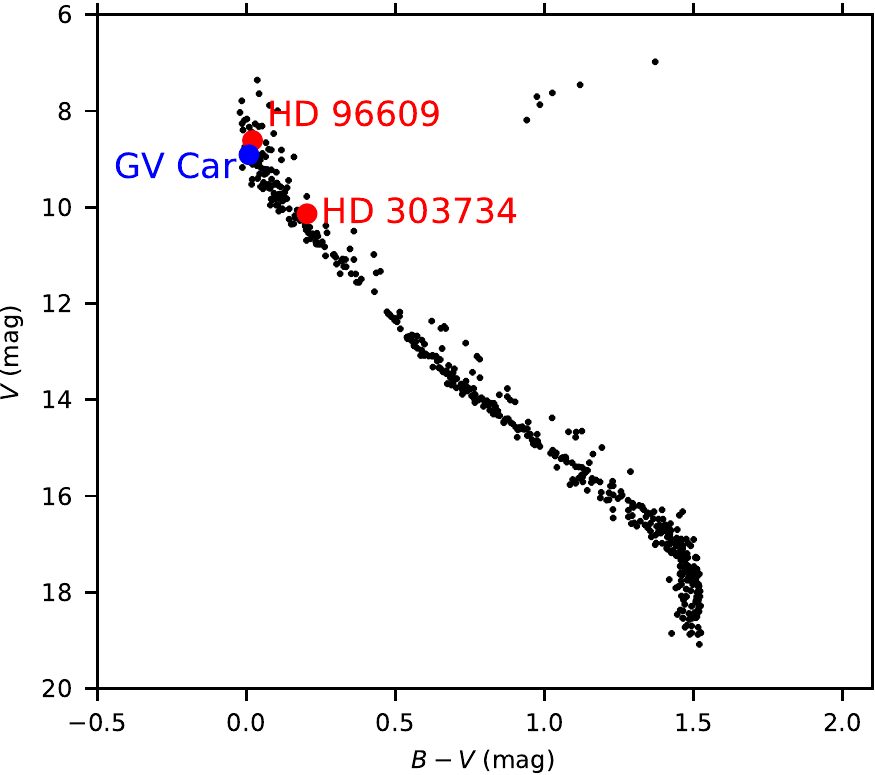}
	\caption{Positions of the target systems and GV\,Car in colour-magnitude diagram of NGC\,3532. 
	Photometric data are taken from \citet{clem_landolt_2011AJ....141..115C}. Plotted data belong 
	to kinematic members of the cluster given in \citet{cantat_gaudin_2018A&A...618A..93C}.}
    \label{fig_cmd}
\end{figure}

Beside photometric studies, \citet{gonzales_lapasette_hd96609_2002AJ....123.3318G} investigated 
spectroscopic binaries and kinematic membership in NGC\,3532. They discovered SB2 nature of HD\,96609, 
which is also a member of the cluster and one of the main targets in this study. Large radial velocity 
semi-amplitudes of the components ($K_{1}=71.23\pm0.25$ \kms and $K_{2}=102.94\pm0.32$ \kms) indicate 
high orbital inclination, i.e. possible eclipsing nature of the system. No eclipse event has been reported 
for this system so far. However, preliminary inspection of space photometry provided by Transiting 
Exoplanet Survey Satellite \citep[TESS,][]{tess_10.1117/1.JATIS.1.1.014003} clearly shows that the system 
is an eclipsing binary with an approximate orbital period of 8.2 day. Further evidences of eclipses is 
noticeable in The All Sky Automated Survey (ASAS) photometry \citep[hereafter, ASAS3;][]{ASAS_Pojmanski_1997, 
ASAS_Pojmanski_2002AcA, ASAS3_Pojmanski_2005AcA}. Preliminary inspection of TESS photometry of the stars 
in the field of NGC\,3532 reveals one more eclipsing binary star, HD\,303734. We show positions of these 
newly identified eclipsing binaries in colour-magnitude diagram of NGC\,3532 in Fig.~\ref{fig_cmd}. 
Both systems are located in the main sequence band of the cluster. Furthermore, we note that HD\,96609 is 
very close to the main sequence turn-off point, thus deserves attention for modelling. 
In Fig.~\ref{fig_cmd}, we also show the position of the first discovered eclipsing binary in the 
region of NGC\,3532, GV\,Car \citep{gvcar_southworth_2006Ap&SS.304..199S}. We note that the positions of 
HD\,96609 and GV\,Car in colour-magnitude diagram, which are very close to each other, is remarkable.

\begin{figure*}
	\includegraphics[scale=1.0]{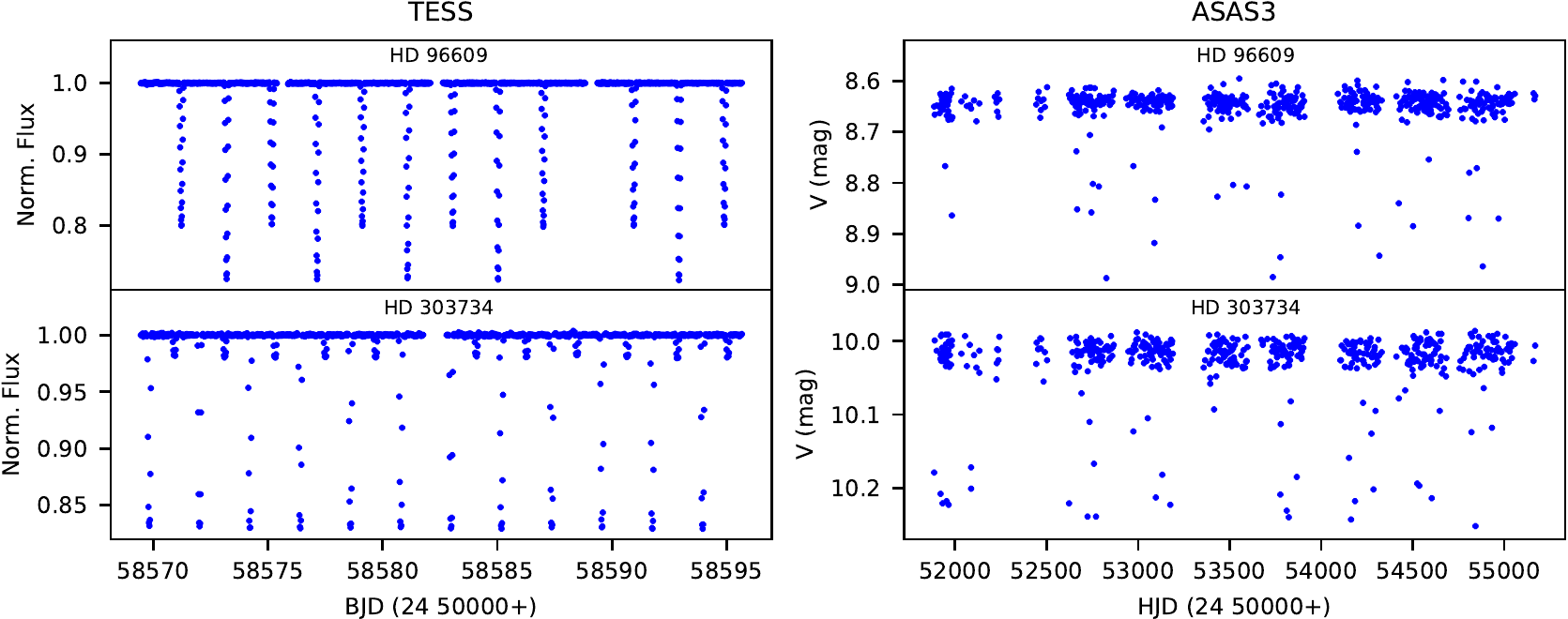}
	\caption{TESS and ASAS3 photometry of the target stars.}
    \label{fig_lc}
\end{figure*}

In this study, we present light curve modelling of newly discovered eclipsing binaries HD\,96609
and HD\,303734. In the case of HD\,96609, we use the advantage of simultaneous analysis of radial
velocity and light curve data to compute physical parameters of the system and its age, which could 
be used to test previous age and distance estimations of NGC\,3532. However, due to the lack of 
spectroscopic data, analysis of HD\,303734 are based on pure photometric data. In the next section, 
we describe the data used in this study, while we give details of light curve and radial velocity 
modelling in Section~\ref{sec_analysis}. In the last section, we summarize and discuss our findings 
with a comparison between physical properties of our target stars and basic properties of NGC\,3532.

\section{Data}\label{sec_data}

Radial velocity measurements of HD\,96609, which are sufficient in number to reveal orbital motion, 
were published by \cite{gonzales_lapasette_hd96609_2002AJ....123.3318G}, where radial velocities 
of both components could be measured. These measurements are based on optical spectra 
recorded by REOSC \'echelle spectrograph with a spectral resolution of 
$R=\lambda/\Delta\lambda=13\,300$. \citet{gonzales_lapasette_hd96609_2002AJ....123.3318G}
reported that the rotational velocities of both components of HD\,96609 is below the 
instrumental error of REOSC spectrograph ($v\,sin\,i<20$ \kms). They obtained radial velocities 
by applying two-dimensional cross-correlation method described in 
\citet{todcor_1994ApJ...420..806Z} via TODCOR algorithm. Target star spectra were cross-correlated 
with spectra of two reference stars (HR\,6041 and HR\,3321). Strong blending in spectral lines of 
both components was observed due to the small radial velocity differences ($v_{r}<35$ \kms) 
between the components in the vicinity of eclipse phases. In these cases, rms residuals are larger 
by factor of $3$ and $4.5$ for the primary and the secondary components, respectively. More details 
on radial velocity measurement technique can be found in \citet{gonzales_lapasette_paperI_2000AJ}. 
We use these measurements for spectroscopic orbit modelling. However, there is no radial velocity 
based orbital solution for HD\,303734, thus analysis of this system purely relies on photometry.

The main source of photometric data is space photometry provided by TESS mission, where Quick Look 
Pipeline (QLP) data from sector $10$ and $11$ observations are available for the target 
systems. Exposure time of these observations is $30$ minutes, which provides pretty good signal-to-noise 
ratio for basic light curve modelling of our targets. We adopt \texttt{KSPSAP FLUX} measurements 
\citep{qlp_tess_huang_2020RNAAS...4..206H}, which are obtained by de-trending simple aperture 
photometry (SAP) fluxes. Before de-trending, high-pass filter are applied to the SAP fluxes in 
order to remove low-frequency variability originating from stellar activity or instrumental noise. 
After de-trending, fluxes are extracted for three apertures with different sizes. Fluxes obtained 
from the best aperture, which is 3 pixels in size, are labelled as \texttt{KSPSAP FLUX}. 
In practice, we obtain \texttt{KSPSAP FLUX} measurements by using \texttt{LIGHTKURVE} package 
written in \texttt{PYTHON} environment \citep{lightkurve_2018ascl.soft12013L}. In the case of crowded 
regions, such as star clusters, extracted fluxes of a given target could be contaminated by 
neighbouring light sources in the close vicinity because of large pixel scale of $21$ arc-seconds 
per pixel for TESS images. However, during extraction of QLP fluxes 
\citet{qlp_tess_huang_2020RNAAS...4..206H} applied an efficient background subtraction algorithm which is 
based on differential photometry of nearby light sources. If the nearby light sources are not varying 
significantly, then the method is capable of removing large part of the background contamination from 
nearby light sources inside 3 pixels of aperture. For a given target, if one or more nearby light sources 
are variable, then the observed light curve amplitude of the main target would be incorrect. 
Since HD\,96609 and HD\,303734 are dominantly bright targets in their separately defined apertures, 
amplitude of possible variability of any background source in the apertures should be negligible, 
thus we assume zero background contribution to the light curves of the target system.

Eclipsing binary natures of these systems are noticeable in ASAS3 photometry obtained in $V$ bandpass. 
ASAS3 data of both systems cover $8$ years of time span starting from HJD 24\,52000 and ending in 
HJD 24\,55000. Observational strategy of ASAS telescopes is to obtain one or a few measurements per 
observing night from a target area in the sky. Depending on orbital period, application of this strategy 
for a sufficiently long time may provide good phase coverage of full orbital cycle for an eclipsing binary, 
including egress, ingress and eclipse phases. We adopt ASAS3 measurements with a quality flag of A or B 
and check for extreme outlier points in the light curve by eye inspection. We show TESS and ASAS3 photometry 
versus barycentric (for TESS data) and heliocentric (for ASAS3 data) Julian Date in Fig.~\ref{fig_lc}.

\section{Analysis}\label{sec_analysis}

We perform light curve modelling with the v40 version of the \textsc{Fortran} code 
\texttt{JKTEBOP}\footnote{https://www.astro.keele.ac.uk/jkt/codes/jktebop.html}
\citep{jktebop1_2004MNRAS.351.1277S, jktebop2_2005A&A...429..645S}, which is mainly 
based on \textsc{Eclipsing Binary Orbit Program} \citep[EBOP,][]{popper_etzel_1981AJ.....86..102P} 
written by Paul Etzel. The program uses biaxial ellipsoidal model 
\citep{nelson_davis_1972ApJ...174..617N} which is also known as Nelson-Davis-Etzel biaxial 
ellipsoidal model. The \texttt{JKTEBOP} code is capable of very fast modelling of well-detached 
eclipsing binaries and equipped with detailed and robust error analysis routines. However, the 
code is not capable of light curve modelling of very close eclipsing binaries since binarity effects due 
to the distorted shapes of the components become dominant at out of eclipse phases and can not be 
fitted by Nelson-Davis-Etzel model properly. It is also not possible to model stellar 
spots and pulsations with the \texttt{JKTEBOP} code.

Inspecting the ASAS3 light curves of HD\,96609, we noticed that there are few data points 
around ingress and egress phases, which may highly reduce the precision of fractional radii resulting 
from the best-fitting model and lead to erroneous results. We observe quantitative evidence of this 
possibility in trial-error light curve modelling process before final modelling stage. In preliminary 
light curve modelling attempts, we observe that resulting fractional radii from the best-fitting models 
for TESS and ASAS3 light curves do not agree. In the case of HD\,96609, more massive star appears as 
the larger star in the system according to the best-fitting light curve model for TESS data, while 
modelling results of ASAS3 data indicate that the same star is the smaller one in the system. 
Furthermore, internal errors of fractional radii are $100$ times larger for ASAS3 model results compared 
to TESS model results. Similar situation is valid for HD\,303734. Moreover, secondary eclipse of 
HD\,303734 is not distinguishable in ASAS3 light curves due to observational scatter. Therefore, 
we exclude ASAS3 data from analyses for both systems.

We have both radial velocity data and high precision TESS photometry for HD\,96609, thus we model 
two data simultaneously. Before modelling, we simply convert normalized TESS fluxes to the magnitudes 
by using the equation $m=-2.5\times\,log(F)$, where $F$ denotes normalized TESS flux. We adjust 
mid-time of the primary eclipse ($T_{0}$), orbital period ($P$), light scale factor ($S$), fractional 
radii of the larger and smaller components ($r_{1}$ and $r_{2}$, respectively), surface brightness 
ratio ($J$), inclination of the orbital plane ($i$), radial velocity semi-amplitudes of the larger 
and smaller components ($K_{1}$ and $K_{2}$, respectively) and center-of-mass velocity of the system 
($V_{\gamma}$). Besides, depending on $K_{1}$ and $K_{2}$, the code internally computes the mass ratio 
of the system $q=M_{2}/M_{1}$, where $M_{1}$ and $M_{2}$ denote the masses of the more massive and less 
massive components, respectively. Since we do not observe any considerable light variation at out-of 
eclipse phases, we fix photometric mass ratio value to a negative number in order to force the stars 
to be spherical. This is a feature of \texttt{JKTEBOP} code. We adopt square root limb darkening 
law \citep{diaz_cordoves_1992A&A...259..227D} for both components of HD\,96609. We use limb darkening 
coefficients of \citet{claret_TESS_limb_darkening_2017A&A...600A..30C} which are computed for TESS 
bandpass by using plane-parallel ATLAS stellar atmosphere models with solar metallicity and $2$\,\kms 
of micro-turbulence velocity. We consider effective temperature and surface gravity of each component 
in order to determine linear ($x$) and non-linear ($y$) limb darkening coefficients by applying linear 
interpolation in the tables of \citet{claret_TESS_limb_darkening_2017A&A...600A..30C}. We do not adjust 
limb darkening coefficients during analyses. Since the model is well constrained by radial velocity 
data and high precision TESS photometry, model convergence is very fast for HD\,96609.

We have TESS photometry but no radial velocity data for HD\,303734. Thus, we only adjust 
$T_{0}$, $P$, $S$, $r_{1}$, $r_{2}$, $J$ and $i$ during light curve modelling of this system. Ignoring 
radial velocity data and related adjustable parameters, applied modelling strategy is the same as in 
the case of HD\,96609. We use square root and linear limb darkening laws for the primary and 
the secondary components, respectively, and we follow the reference and strategy mentioned above for 
determination of limb darkening coefficients. 

\begin{figure}
	\includegraphics[scale=0.73]{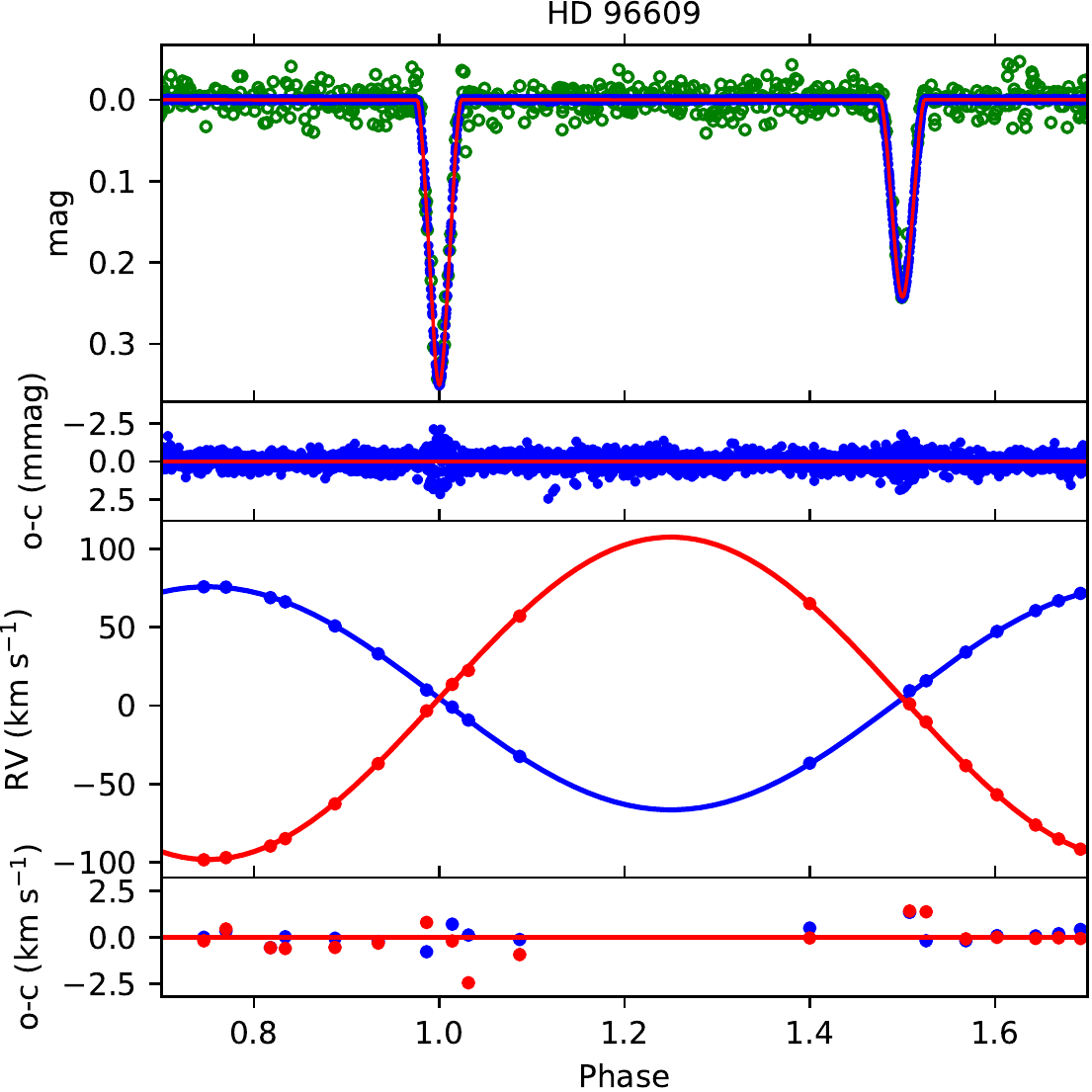}
	\caption{TESS photometry (blue points), radial velocity measurements (blue and red points for 
	the more massive and less massive components, respectively) and the best-fitting light curve 
	and spectroscopic orbit models of HD\,96609 (continuous curves). We also over plot ASAS3 photometry 
	(green open circles) for comparison. Note that we apply constant shift of 8\fm64 to ASAS3 magnitudes 
	in order to evaluate TESS and ASAS3 data in a common scale.}
    \label{fig_hd96609_model}
\end{figure}

\begin{figure}
	\includegraphics[scale=0.73]{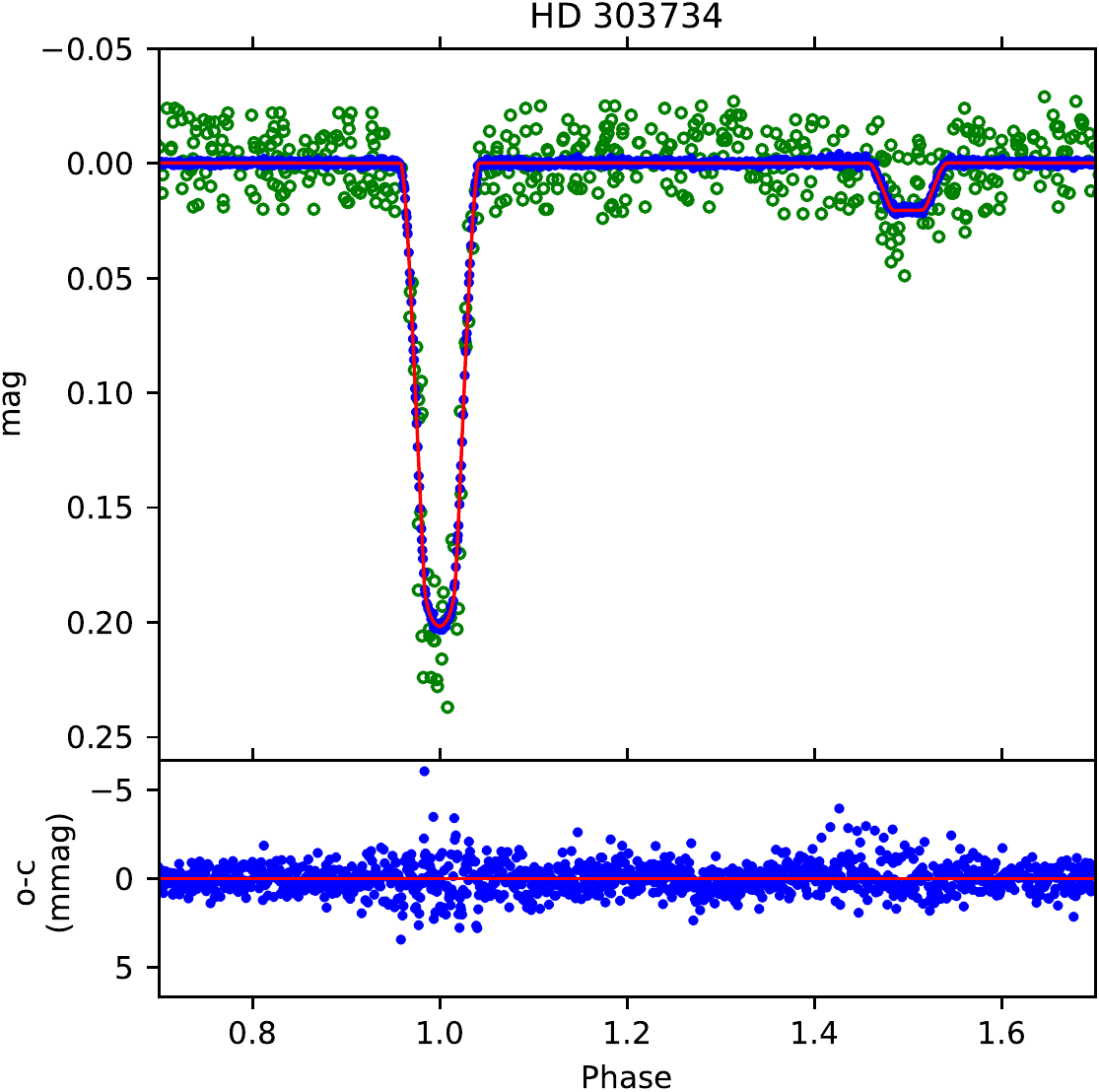}
	\caption{TESS photometry (blue points) and the best-fitting light curve model of HD\,303734 
	(continuous curve). ASAS3 photometry is shown by green open circles for comparison. Note that 
	we apply constant shift of 10\fm015 to ASAS3 magnitudes in order to evaluate TESS and ASAS3 
	data in a common scale.}
    \label{fig_hd303734_model}
\end{figure}

We show phase-folded observations and the best-fitting models in Fig.~\ref{fig_hd96609_model} and 
\ref{fig_hd303734_model} for HD\,96609 and HD\,303734, respectively. We note that we show ASAS3 
data only for comparison purposes in these figures. Although the best-fitting models for observational 
data of each system appear fairly good in these figures, we discuss stability of the best-fitting model 
for HD\,303734 in later paragraphs.

\begin{figure*}
	\includegraphics[scale=0.32]{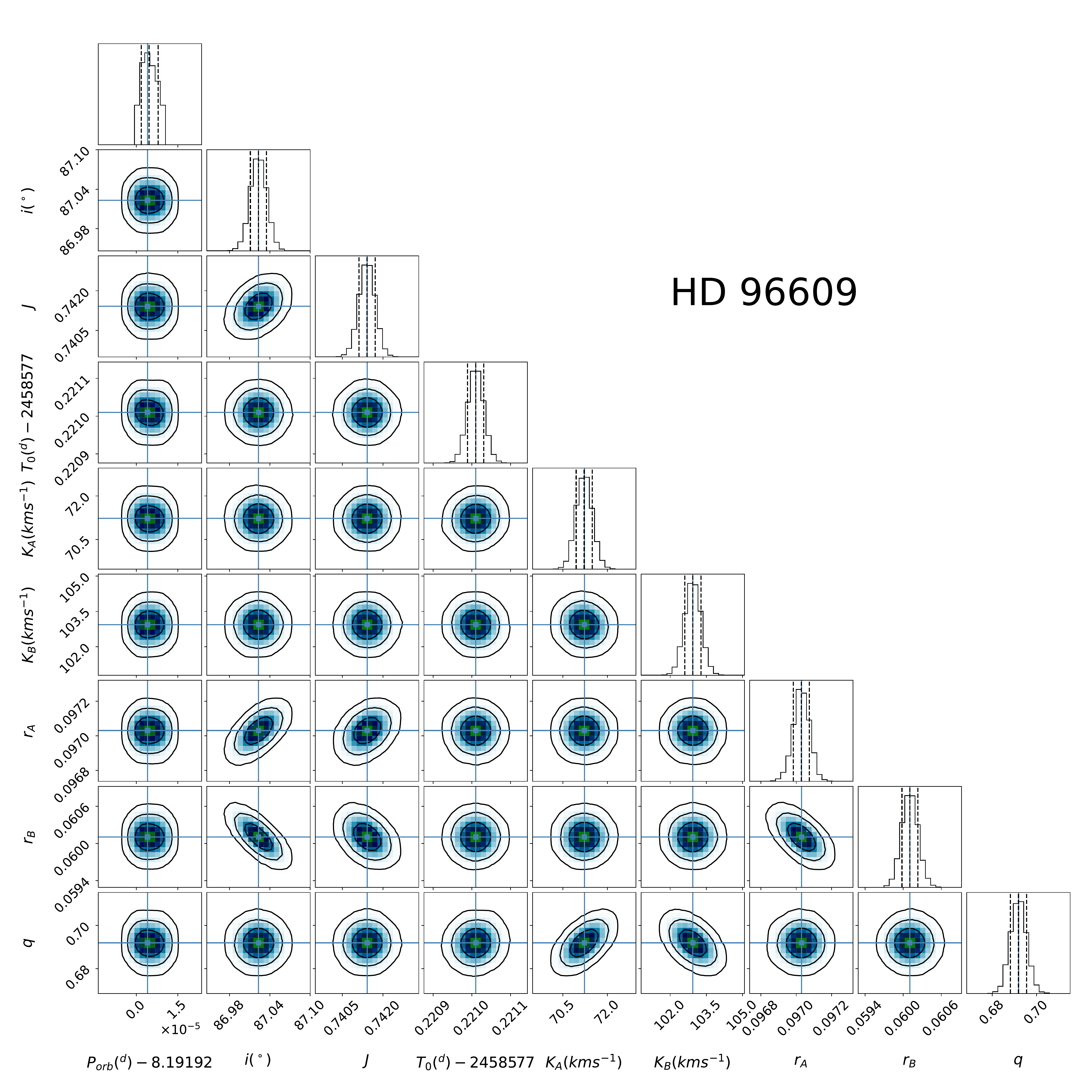}
	\caption{Distribution of results from 10\,000 Monte Carlo simulations for HD\,96609. 
	The figure is made by using \texttt{corner.py} script \citep{corner_plot}. Three 
	contour levels in each plot window show 1, 2 and 3$\sigma$ levels. Diagonal histogram
	plots show posterior distribution of each parameter.}
    \label{fig_corner_HD96609}
\end{figure*}

\begin{figure}
	\includegraphics[scale=0.23]{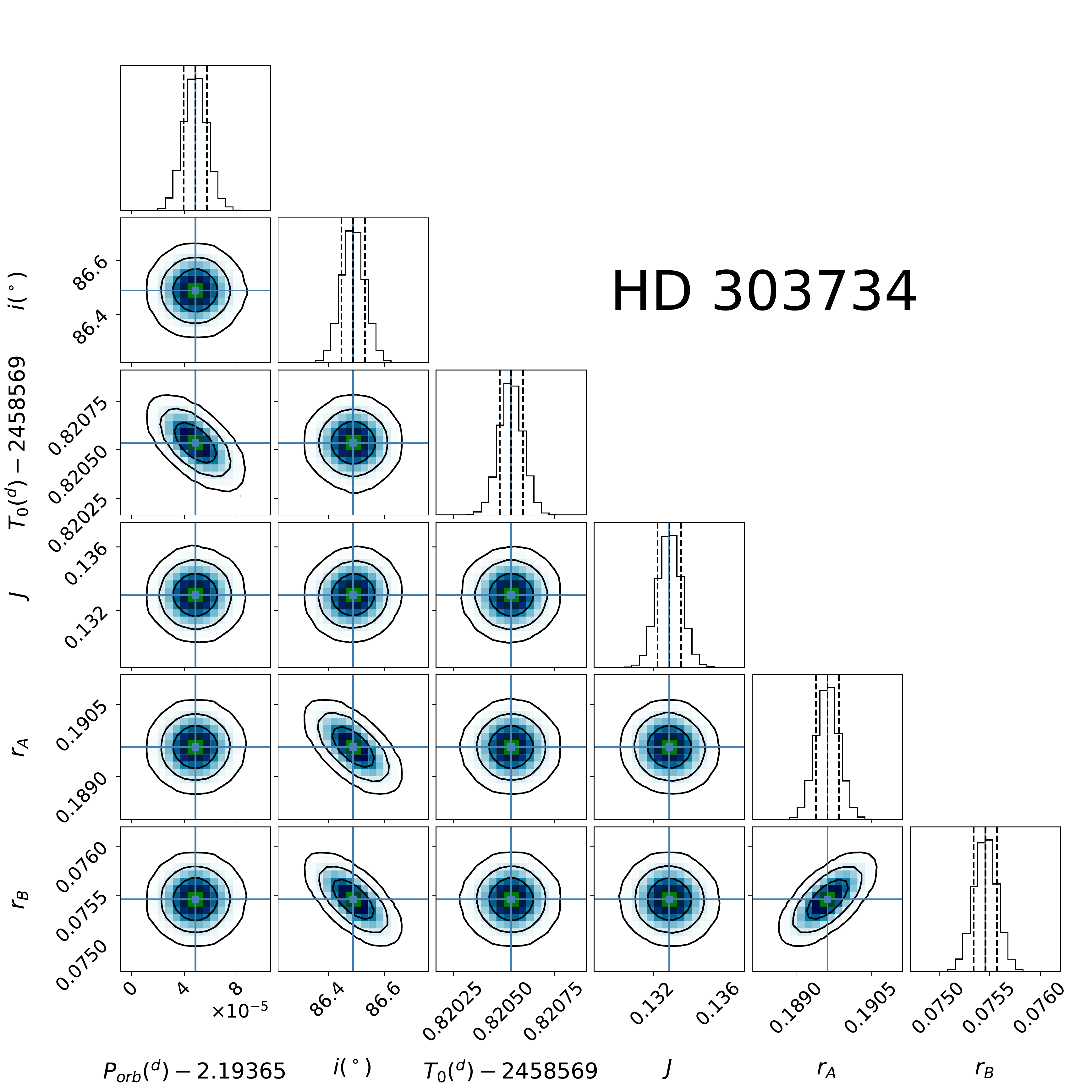}
	\caption{Distribution of results from 10\,000 Monte Carlo simulations for HD\,303734.}
    \label{fig_corner_HD303734}
\end{figure}

One may obtain formal uncertainties of final model parameters from solution covariance matrix. \texttt{JKTEBOP} 
code provides these uncertainties at the end of each solution run. However, these uncertainties are 
mostly underestimated or unrealistically small due to the possible strong correlations between adjusted 
parameters. In order to compute more realistic uncertainties, we use TASK8 feature of the 
\texttt{JKTEBOP}, which applies Monte Carlo simulations. In application of TASK8 for each target, the
code re-evaluates the best-fitting model at the phases of the actual observations by adding Gaussian
noise. We set the code to repeat this process for 10\,000 times and compute a set of solution parameters
for each process. Finally, we compute standard deviation of the 10\,000 different results of each 
adjustable parameter and adopt this standard deviation as the final uncertainty of the corresponding 
adjusted parameter. We tabulate final best-fitting model parameters and 1$\sigma$ uncertainties 
estimated from Monte Carlo simulations in Table~\ref{table_results}. We note that the lack of radial 
velocity observations between 1.1 and 1.4 phases might probably responsible for a significant amount 
of the uncertainties tabulated for $K_{1}$ and $K_{2}$ parameters in the table. We also show 
distribution of Monte Carlo simulations for each target in Fig.~\ref{fig_corner_HD96609} and 
\ref{fig_corner_HD303734}, in order to illustrate correlations between adjusted parameters. 

Inspecting Fig.~\ref{fig_corner_HD96609}, we do not see any remarkable skewness in correlation plots, 
which is clearly the result of good parameter constraints provided by simultaneous modelling of light 
curve and radial velocity data. On the other hand, model parameter constraint is not very good due to 
the absence of spectroscopic data in the case of HD\,303734. Slight skewness in correlation plots of 
some parameters may easily be noticed in Fig.~\ref{fig_corner_HD303734}. However, we are still able to
obtain reasonable representation of TESS light curve of HD\,303734.

\begin{table}
	\centering
	\caption{\texttt{JKTEBOP} modelling results of the target stars. 
	In the lower part of the table, we tabulate computed absolute parameters of 
	HD\,96609.}
	\label{table_results}
	\begin{tabular}{lcc} 
		\hline
Model Parameters	&	HD\,96609	&	HD\,303734 \\
\hline\noalign{\smallskip}
$T_{0}$ (BJD)	&	$58577.22101\,(2)$	&	$58569.82053\,(5)$	\\
$P$ (day)	&	$8.191924\,(3)$	&	$2.193648\,(9$)	\\
$S$ (mag)	&	$-0.000040\,(8)$	&	$-0.00016\,(2)$	\\
$J$	&	$0.7414\,(3)$	&	$0.1329\,(7)$	\\
$x_{1},\,y_{1}$	&	$-0.038$, $0.553$	&	$-0.072$, $0.668$	\\
$x_{2},\,y_{2}$	&	$-0.027$, $0.561$	&	$0.559$, ---	\\
$r_{1}$	&	$0.09703\,(4)$	&	$0.1896\,(2)$	\\
$r_{2}$	&	$0.00010\,(12)$	&	$0.0754\,(1)$	\\
$i~(^{\circ})$	&	$87.02\,(1)$	&	$86.48\,(4)$	\\
$K_{1}$ (\kms)	&	$71.2\,(3)$	&	---	\\
$K_{2}$ (\kms)	&	$102.9\,(3)$	&	---	\\
$V_{\gamma}$ (\kms)	&	$4.6\,(1)$	&	---	\\
$e$	&	$0$	&	$0$	\\
$\omega~(^{\circ})$	&	---	&	---	\\
$q=M_{2}/M_{1}$	&	$0.692\,(4)$	&	---	\\
lc rms (mmag)	&	$0.455$	&	$0.834$	\\
$rv_{1}$ rms (\kms)	&	$0.46$	&	---	\\
$rv_{2}$ rms (\kms)	&	$0.84$	&	---	\\
	&		&		\\
Physical Parameters	&		&		\\
$a$ (\Rsun)	&	$28.23\,(6)$	&	---	\\
$M_{1}$ (\Msun)	&	$2.66\,(2)$	&	---	\\
$M_{2}$ (\Msun)	&	$1.84\,(1)$	&	---	\\
$R_{1}$ (\Rsun)	&	$2.740\,(6)$	&	---	\\
$R_{2}$ (\Rsun)	&	$1.697\,(5)$	&	---	\\
$log\,g_{1}$ (cgs)	&	$3.987\,(1)$	&	---	\\
$log\,g_{2}$ (cgs)	&	$4.243\,(2)$	&	---	\\

$T_{1}$ (K)	&	$9650\,(250)$	&	$7970\,(500)$	\\
$T_{2}$ (K)	&	$8950\,(250)$	&	$4812\,(300)$	\\
$log\,L_{1}$ (\Lsun)	&	$1.76\,(4)$	&	---	\\
$log\,L_{2}$ (\Lsun)	&	$1.22\,(4)$	&	---	\\
$M_{Bol_{1}}$ (mag)	&	$0.32\,(11)$	&	---	\\
$M_{Bol_{2}}$ (mag)	&	$1.69\,(12)$	&	---	\\
$d$ (pc)	&	$460\,(17)$	&	---	\\
		\hline

	\end{tabular}
\end{table}

Since \texttt{JKTEBOP} does not work with effective temperatures, we can not directly include 
effective temperatures in modelling process. However, we may use photometric data of HD\,96609 
as an auxiliary property and estimate the effective temperatures, which enables us to compute 
the distance of the system. Adopting $E(B-V)=0\fm028$ and $B-V=0\fm021\pm0\fm010$ from 
\citet{clem_landolt_2011AJ....141..115C}, we find $(B-V)_{0}=-0\fm007\pm0\fm010$ corresponding 
to an effective temperature of $9650$ K and spectral type of B9-A0V according to the calibrations 
given by \citet{gray_2005}. Uncertainty of $B-V$ colour indicates $120$ K of uncertainty on the 
effective temperature according to this calibration. However, $120$ K of uncertainty estimated 
from broad band photometry is probably underestimated for such a hot 
star, thus we believe that the real uncertainty is higher and approximately around $250$ K. 
Assuming that the resulting surface brightness ratio $J$ in Table~\ref{table_results} could be 
considered as approximate ratio of the effective temperatures in terms of $(T_{2}/T_{1})^{4}$,
we compute the effective temperature of the secondary component as $8950$ K, corresponding to A2V spectral 
type. Although $J$ is computed via photometric data obtained in TESS bandpass, it is still fairly reasonable
to assume that $J\approx(T_{2}/T_{1})^{4}$ since temperature difference between the primary and the secondary 
component is not large. Feeding these values, adopted $V$ magnitude and $E(B-V)$ 
values of HD\,96609 from \citet{clem_landolt_2011AJ....141..115C} together with $P$, $K_{1}$, $K_{2}$, $i$, $r_{1}$, $r_{2}$, 
$T_{1}$ and $T_{2}$,into the \texttt{JKTABSDIM}\footnote{https://www.astro.keele.ac.uk/jkt/codes/jktabsdim.html} 
code, we obtain absolute physical properties and the distance of HD\,96609. \texttt{JKTABSDIM} code
computes absolute bolometric magnitudes of the components by assuming $T_{eff}=5780$ K and $M_{bol}=4\fm74$ 
as solar values. Considering various calibrations \citep{bessell_1998A&A...333..231B,girardi_2002A&A...391..195G,
code_1976ApJ...203..417C, flower_1996ApJ...469..355F, kervella_2004A&A...426..297K} the code applies bolometric 
correction in $V$ band to the computed absolute bolometric magnitudes, which ultimately provides absolute $V$ 
magnitudes of the components separately, overall absolute $V$ magnitude of the system and finally the distance of 
the system for each calibration separately. Averaging distance values from calibrations referred above, we finally 
compute the average distance of the system. We tabulate all results in the second column of Table~\ref{table_results}.

Light curve modelling of HD\,303734 reveals that the system is a totally eclipsing binary.
Although lack of radial velocity data prevents us from computing precise physical parameters,
we may use photometric properties of the system to reveal its nature a little more. Following the
same method proposed for HD\,96609 above and adopting same calibrations, we find 
$(B-V)_{0}=0\fm176\pm0\fm041$ corresponding to an average effective temperature of $7970\pm500$ K
and spectral type of A6V. Main source of larger uncertainty compared to the case of HD\,96609 is the 
$\sim6$ times larger uncertainty on $V$ measurement \citep[$V=10\fm135\pm0\fm040$,][]{clem_landolt_2011AJ....141..115C} 
compared to the $V$ measurement of HD\,96609. Surface brightness ratio of the 
components found in the light curve modelling suggests $4812$ K for the effective temperature of the 
secondary component, indicating K3V spectral type. Large temperature difference between the components 
of HD 303734 is remarkable. In this case, effective temperature estimation of the secondary component via 
$J\approx(T_{2}/T_{1})^{4}$ approximation can not be very precise because the surface brightness ratio depends 
on not only the ratio of the temperatures but also the bandpass used in the photometry. In the case of HD\,303734,
components of the system appear to possess different spectral energy distribution, thus $J$ would be very likely
different in different specific wavelength ranges (i.e. bandpasses). Therefore, estimated
$4812$ K effective temperature of the secondary component should be considered with caution. Due to the lack of 
spectroscopic orbit solution, we refrain from further evaluation of the system since indirect estimations of physical 
properties would include very large uncertainties.


\section{Discussion}\label{sec_discussion}

\begin{figure}
	\includegraphics[scale=0.73]{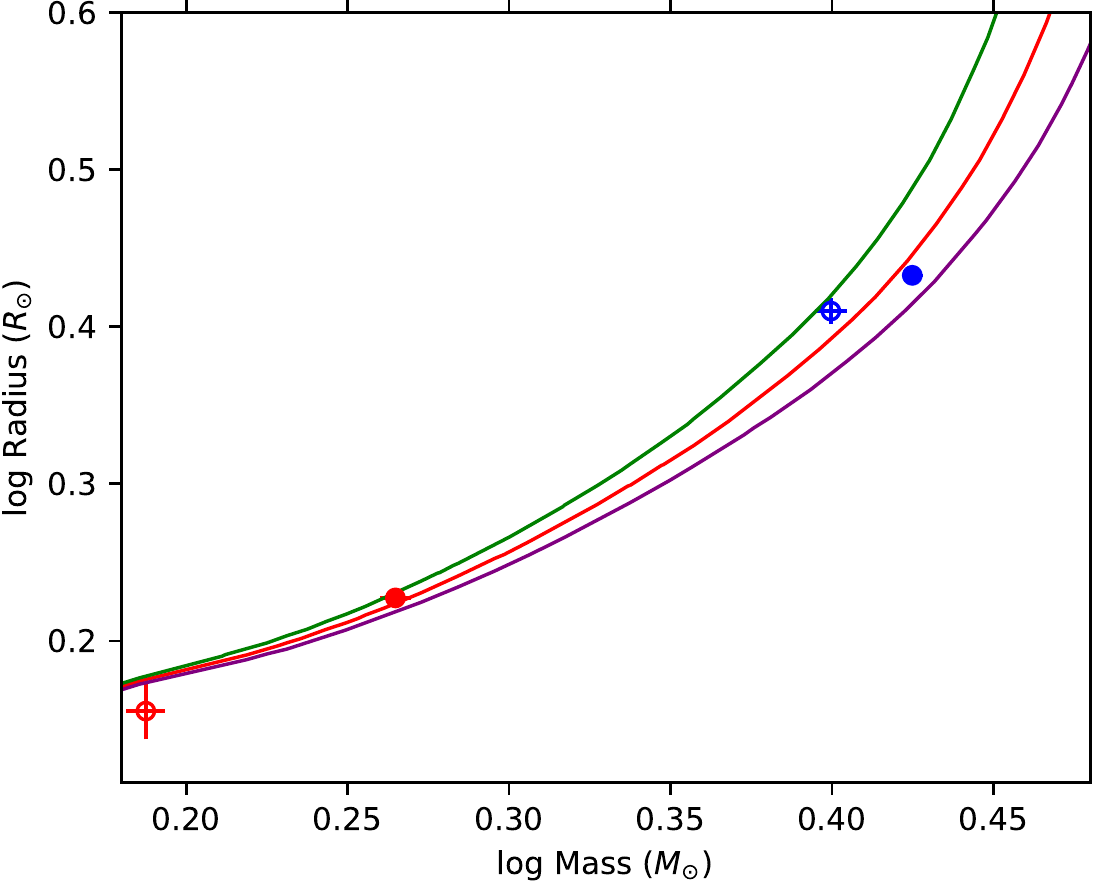}\\
	\includegraphics[scale=0.73]{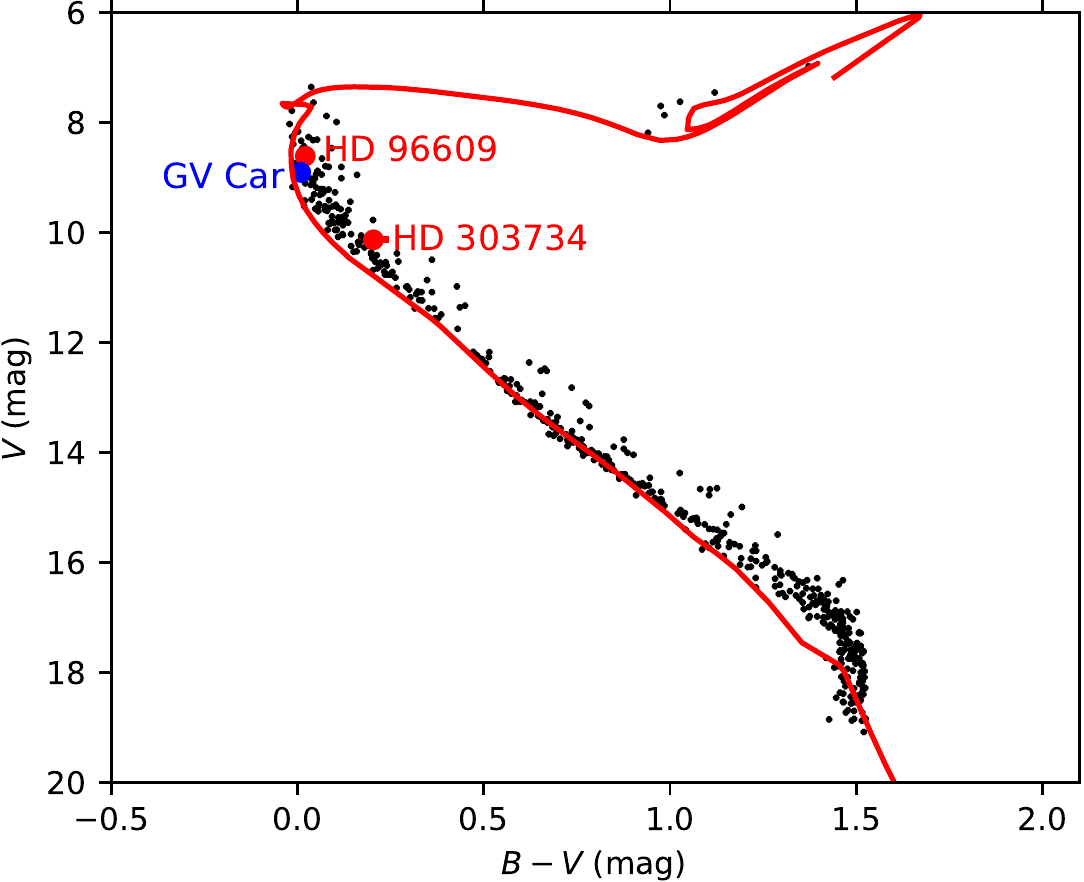}
	\caption{Upper panel: Positions of the components of HD\,96609 on $log~M-log~R$ plane 
	(filled circles). Both components of GV\,Car are also plotted in the panel with open circles.
	Plotted isochrones are for log(age/yr) 8.5 (purple), 8.55 (red) and 8.6 (green). Note that error bars
	are mostly smaller than the size of filled circles for the components fo HD\,96609. 
	Lower panel: Same as Figure~\ref{fig_cmd} but with log(age/yr) 8.55 isochrone over plotted.}
    \label{fig_mr_plane}
\end{figure}

Simultaneous light curve and spectroscopic orbit modelling of HD\,96609 enables us to compute 
precise masses and radii of the components. We plot components of HD\,96609 on $log~M-log~R$ plane 
in upper panel of Fig.~\ref{fig_mr_plane} together with Precomputed PAdova and TRieste Stellar 
Evolution Code (PARSEC) isochrones \citet{Bressan_et_al_2012MNRAS.427..127B} for solar metallicity 
with $Y=0.279$ and 	$Z=0.017$. Inspecting positions of the components, we find that the 
best-fitting isochrone has log(age/yr) 8.55. Considering uncertainty of masses in the figure and 
plotted isochrones, we estimate the age of the system as $350\pm40$ Myr, which nicely agrees 
with the $300\pm100$ Myr age estimation of \citet{clem_landolt_2011AJ....141..115C}. In the figure, 
we also over plot the components of the detached eclipsing binary GV\,Car, which is 
also the member of the cluster and the first eclipsing binary discovered in NGC\,3532 
\citep{gvcar_southworth_2006Ap&SS.304..199S}. Position of the primary component of GV\,Car is 
consistent with the isochrone of log(age/yr) 8.55 but the secondary component appears outside of the
plotted isochrones, even if we consider the $1\sigma$ error bars. Eclipse depth variation of GV\,Car 
was reported in the same study and two possible explanations suggested for this variation. There is
either a decrease in the orbital inclination by 3$^{\circ}$ due to a perturbed orbit or an increase 
in brightness of a possible third light. Thus, the positional inconsistency of the secondary 
component of GV\,Car in Fig.~\ref{fig_mr_plane} might partly be related to this yet fully unexplained 
nature of the system. Over plotting log(age/yr) 8.55 isochrone in colour-magnitude diagram of the cluster 
(lower panel of Fig.~\ref{fig_mr_plane}), we see reasonable agreement with the observed data, which
supports $350\pm40$ Myr of age. Isochrones plotted in Fig.~\ref{fig_mr_plane} also confirm that the 
cluster possesses near solar metallicity.

Computed $460\pm17$ pc distance of HD\,96609 in this study agrees with the $484^{+35}_{-30}$ 
pc distance computed from GAIA parallaxes within $1\sigma$ error. However, considering $1\sigma$ 
error, our distance estimation does not agree with the $492^{+12}_{-11}$ pc distance given by 
\citet{clem_landolt_2011AJ....141..115C}. Comparing three distance estimations, one may
note that the distance estimation of \citet{clem_landolt_2011AJ....141..115C} is quite close
to the distance estimation based on GAIA parallaxes, while the distance we compute is
lower than those estimates. Here, we note that our distance estimation is very sensitive to our
effective temperature estimation for the primary component, which is based on de-reddened $B-V$ 
colour, and the secondary component, which is computed via surface brightness ratio found from the 
light curve modelling. Further spectroscopy of the system may help to confirm or improve effective 
temperature estimations, thus the distance of the system. On the other hand, it is put forwarded
that there are strong indications that GAIA parallaxes are affected from a small bias 
and these parallaxes should be smaller \citep{gaia_bias_Lindegren_2021A&A...649A...4L, 
gaia_parallaxes_groenewegen_2021arXiv210608128G}. This might partly explain the difference between 
computed distance in this study and the estimated distance of the cluster based on GAIA parallaxes.

We are only able to reveal basic light curve model properties of HD\,303734 without its 
physical properties. Best-fitting light curve model and photometric properties of the system indicate
that HD\,303734 is a totally eclipsing binary composed of a hot primary component possessing A6V 
spectral type \citep[w.r.t its estimated effective temperature,][]{gray_2005} and very cool (likely 
K3V spectral type) secondary component. Modelling results suggest that both components of the system 
are main sequence stars. However, model parameter constraint is not as good as in the case of HD\,96609.
Future optical spectroscopic observations of this system may provide further constraints for a more precise
model and test for our findings in this study. Then, it would be possible to evaluate this system as 
additional reference for further tests on the age and the distance of NGC\,3532.

\section*{Acknowledgements}
I express my thanks to Bar\i\c{s} Hoyman for his help on production of some figures in this study. 
I would like to thank anonymous referee for his/her thoughtful comments and critically reading, which 
improve and clarify the manuscript. This paper includes data collected by the TESS mission, which are 
publicly available from the Mikulski Archive for Space Telescopes (MAST). Funding for the TESS mission 
is provided by the NASA's Science Mission Directorate. This research has made use of NASA's Astrophysics 
Data System. This research has made use of the SIMBAD database, operated at CDS, Strasbourg, France.

\textit{Software}: PYTHON, NUMPY \citep{numpy_harris_2020array}, MATPLOTLIB \citep{matplotlib_Hunter:2007}, 
SCIPY \citep{scipy_2020_NMeth}.

\section*{DATA AVAILABILITY}
ASAS V-band photometric data of HD\,96609 available from \url{http://www.astrouw.edu.pl/cgi-asas/asas_cgi_get_data?110327-5829.8,asas3}.
ASAS V-band photometric data of HD\,303734 available from \url{http://www.astrouw.edu.pl/cgi-asas/asas_cgi_get_data?110651-5842.4,asas3}.
The TESS QLP data used in this paper are available on MAST.

\bibliographystyle{mnras}
\bibliography{ngc3532_ebs} 





\bsp	
\label{lastpage}
\end{document}